\documentclass[aps, prb, twocolumn, superscriptaddress,preprintnumbers]{revtex4-2}
\usepackage{bm, amsmath, amsfonts, amssymb, ascmac, mathtools, braket}
\usepackage{times}
\usepackage{multirow}
\usepackage{graphicx}
\usepackage{float, color, xcolor}

\usepackage[whole]{bxcjkjatype} % for Japanese
\usepackage{subfigure} % for subfigure
\usepackage{amscd} % for homological alg

\usepackage{bbm}
\usepackage{tabularx}

\usepackage{comment}

\usepackage[
pagebackref=false,
colorlinks=true,
linkcolor=blue,
urlcolor=blue,
filecolor=black,
citecolor=red,
pdfstartview=FitV,
pdftitle={},
pdfauthor={},
pdfsubject={},
pdfkeywords={},
pdfpagemode=None,
bookmarksopen=true
]{hyperref}

\newcommand{\ii}{\text{i}}

\newcommand{\bk}{{\bm{k}}}

\newcommand{\tr}{{{\rm tr\ }}}
\newcommand{\Z}{{\mathbb{Z}}}

\begin{document}

\title{$K$-theory classification of Wannier localizability and detachable topological boundary states}

\author{Ken Shiozaki}
\email{ken.shiozaki@yukawa.kyoto-u.ac.jp}
\affiliation{Center for Gravitational Physics and Quantum Information, Yukawa Institute for Theoretical Physics, Kyoto University, Kyoto 606-8502, Japan}

\author{Daichi Nakamura}
\email{daichi.nakamura@issp.u-tokyo.ac.jp}
\affiliation{Institute for Solid State Physics, University of Tokyo, Kashiwa, Chiba 277-8581, Japan}

\author{Kenji Shimomura}
%\email{kenji.shimomura@yukawa.kyoto-u.ac.jp}
\affiliation{Center for Gravitational Physics and Quantum Information, Yukawa Institute for Theoretical Physics, Kyoto University, Kyoto 606-8502, Japan}

\author{Masatoshi Sato}
%\email{msato@yukawa.kyoto-u.ac.jp}
\affiliation{Center for Gravitational Physics and Quantum Information, Yukawa Institute for Theoretical Physics, Kyoto University, Kyoto 606-8502, Japan}

\author{Kohei Kawabata}
\email{kawabata@issp.u-tokyo.ac.jp}
\affiliation{Institute for Solid State Physics, University of Tokyo, Kashiwa, Chiba 277-8581, Japan}

\date{\today}
\preprint{YITP-24-88}

\begin{abstract}
A hallmark of certain topology, including the Chern number, is the obstruction to constructing exponentially localized Wannier functions in the bulk bands.
Conversely, other types of topology do not necessarily impose the Wannier obstructions.
Remarkably, such Wannier-localizable topological insulators can host boundary states that are detachable from the bulk bands.
In our accompanying Letter [D. Nakamura {\it et al.}, Phys. Rev. Lett. $\bm{135}$, 096601 (2025), arXiv:2407.09458], we demonstrate that non-Hermitian topology underlies detachable boundary states in Hermitian topological insulators and superconductors, thereby establishing their tenfold classification based on internal symmetry.
Here, using $K$-theory, we elucidate the relationship between Wannier localizability and detachability of topological boundary states.
From the boundary perspective, we classify intrinsic and extrinsic non-Hermitian topology, corresponding to nondetachable and detachable topological boundary states, respectively.
From the bulk perspective, on the other hand, we classify Wannier localizability through the homomorphisms of topological phases from the tenfold Altland-Zirnbauer symmetry classes to the threefold Wigner-Dyson symmetry classes.
These two approaches from the boundary and bulk perspectives lead to the same classification.
We clarify this agreement and develop a unified understanding of the bulk-boundary correspondence on the basis of $K$-theory.
\end{abstract}

\maketitle

%%%%% Introduction %%%%%
\section{Introduction}

Topological phases of matter lay a foundation in modern condensed matter physics~\cite{HK-review, QZ-review}.
Specifically, topological phases of band insulators and gapped Bogoliubov-de Gennes superconductors are generally classified within the fundamental tenfold internal symmetry classification based on time reversal, particle-hole transformation, and chiral transformation~\cite{AZ-97, Schnyder-08, *Ryu-10, Kitaev-09, CTSR-review}.
A hallmark of the topologically nontrivial bulk bands is the emergence of anomalous boundary states.
Furthermore, certain topology, such as the Chern number and the time-reversal-invariant $\mathbb{Z}_2$ topological invariant, imposes obstructions to constructing exponentially localized Wannier functions in the bulk bands~\cite{Thouless-84, Brouder-07, Soluyanov-11, Read-17, Vanderbilt-textbook}.
The Wannier localizability also refines definition of topological crystalline phases~\cite{Po-18}.
By contrast, other types of topology do not necessarily impose the Wannier obstructions.
Examples include one-dimensional topological insulators~\cite{Kohn-59, Kivelson-82}, three-dimensional topological superconductors~\cite{ono-20}, and three-dimensional topological insulators protected by chiral symmetry~\cite{Altland-24}.
Remarkably, these Wannier-localizable topological insulators do not necessitate spectral flow of the concomitant boundary states.
As opposed to the conventional intuition of the bulk-boundary correspondence, such topological boundary states are detachable from the bulk bands~\cite{Alexandradinata-21, Altland-24}.

In our accompanying Letter~\cite{NSSSK-24}, we elucidate that non-Hermitian topology underlies Wannier localizability and detachable boundary states in Hermitian topological insulators and superconductors, thereby establishing their tenfold classification (Table~\ref{tab: classification}).
This connection is based on the identification of topological boundary states as non-Hermitian topology~\cite{JYLee-19, Bessho-21, Ma-24, Schindler-23, Nakamura-23, Hamanaka-24}.
In general, complex-valued spectra of non-Hermitian systems allow two types of complex-energy gaps: point and line gaps~\cite{Gong-18, *Kawabata-19, KSUS-19}.
In the presence of a point (line) gap, complex-energy bands are defined not to cross a reference point (line) in the complex-energy plane.
Non-Hermitian systems with point gaps are continuously deformable to unitary systems and can thus be intrinsic to non-Hermitian systems~\cite{OKSS-20, Shiozaki}.
Conversely, non-Hermitian systems with real (imaginary) line gaps are continuously deformable to Hermitian (anti-Hermitian) systems and hence have Hermitian counterparts.
We demonstrate that intrinsic non-Hermitian topology prohibits the detachment of topological boundary states, whereas extrinsic non-Hermitian topology enables their detachment and further leads to anti-Hermitian topology of the detached boundary states.
Building on this connection, we also establish the tenfold classification of Wannier localizability and detachable topological boundary states, as well as topology of the detached boundary states.
We also note that Ref.~\cite{Altland-24} also provided the classification of Wannier localizability for one, two, and three dimensions.

\begin{table*}[t]
	\centering
	\caption{Wannier localizability of the $d$-dimensional bulk and detachability of $\left( d-1 \right)$-dimensional boundary states in topological insulators and superconductors.
    The tenfold Altland-Zirnbauer symmetry classes consist of time-reversal symmetry (TRS), particle-hole symmetry (PHS), and chiral symmetry (CS).
    The entries with ``$\checkmark$" (``$\times$") accompany the Wannier nonlocalizable (localizable) topological phases in the bulk and exhibit nondetachable (detachable) boundary states with intrinsic (extrinsic) non-Hermitian topology.
    The entries with ``$\checkmark/\times$" specify the Wannier nonlocalizable and localizable topological phases for the odd and even numbers of topological invariants, respectively.}
	\label{tab: classification}
     \begin{tabular}{c|ccc|cccccccc} \hline \hline
    ~~Class~~ & ~~TRS~~ & ~~PHS~~ & ~~CS~~ & ~~$d=1$~~ & ~~$d=2$~~ & ~~$d=3$~~ & ~~$d=4$~~ & ~~$d=5$~~ & ~~$d=6$~~ & ~~$d=7$~~ & ~~$d=8$~~ \\ \hline
    A & $0$ & $0$ & $0$ & $0$ & $\mathbb{Z}^{\checkmark}$ & $0$ & $\mathbb{Z}^{\checkmark}$ & $0$ & $\mathbb{Z}^{\checkmark}$ & $0$ & $\mathbb{Z}^{\checkmark}$ \\
    AIII & $0$ & $0$ & $1$ & $\mathbb{Z}^{\times}$ & $0$ & $\mathbb{Z}^{\times}$ & $0$ & $\mathbb{Z}^{\times}$ & $0$ & $\mathbb{Z}^{\times}$ & $0$ \\ \hline
    AI & $+1$ & $0$ & $0$ & $0$ & $0$ & $0$ & $2\mathbb{Z}^{\checkmark}$ & $0$ & $\mathbb{Z}_2^{\checkmark}$ & $\mathbb{Z}_2^{\checkmark}$ & $\mathbb{Z}^{\checkmark}$ \\
    BDI & $+1$ & $+1$ & $1$ & $\mathbb{Z}^{\times}$ & $0$ & $0$ & $0$ & $2\mathbb{Z}^{\times}$ & $0$ & $\mathbb{Z}_2^{\times}$ & $\mathbb{Z}_2^{\times}$ \\
    D & $0$ & $+1$ & $0$ & $\mathbb{Z}_2^{\times}$ & $\mathbb{Z}^{\checkmark}$ & $0$ & $0$ & $0$ & $2\mathbb{Z}^{\checkmark}$ & $0$ & $\mathbb{Z}_2^{\times}$ \\
    DIII & $-1$ & $+1$ & $1$ & $\mathbb{Z}_2^{\times}$ & $\mathbb{Z}_2^{\checkmark}$ & $\mathbb{Z}^{\checkmark/\times}$ & $0$ & $0$ & $0$ & $2\mathbb{Z}^{\times}$ & $0$ \\
    AII & $-1$ & $0$ & $0$ & $0$ & $\mathbb{Z}_2^{\checkmark}$ & $\mathbb{Z}_2^{\checkmark}$ & $\mathbb{Z}^{\checkmark}$ & $0$ & $0$ & $0$ & $2\mathbb{Z}^{\checkmark}$ \\
    CII & $-1$ & $-1$ & $1$ & $2\mathbb{Z}^{\times}$ & $0$ & $\mathbb{Z}_2^{\times}$ & $\mathbb{Z}_2^{\times}$ & $\mathbb{Z}^{\times}$ & $0$ & $0$ & $0$ \\
    C & $0$ & $-1$ & $0$ & $0$ & $2\mathbb{Z}^{\checkmark}$ & $0$ & $\mathbb{Z}_2^{\times}$ & $\mathbb{Z}_2^{\times}$ & $\mathbb{Z}^{\checkmark}$ & $0$ & $0$ \\
    CI & $+1$ & $-1$ & $1$ & $0$ & $0$ & $2\mathbb{Z}^{\times}$ & $0$ & $\mathbb{Z}_2^{\times}$ & $\mathbb{Z}_2^{\checkmark}$ & $\mathbb{Z}^{\checkmark/\times}$ & $0$ \\ \hline \hline
  \end{tabular}
\end{table*}

In the present work, we develop the $K$-theory classification, further elucidating the relationship between Wannier localizability and detachability of topological boundary states.
On the one hand, we classify intrinsic and extrinsic non-Hermitian topology, corresponding to nondetachable and detachable boundary states in Hermitian topological materials, respectively.
On the other hand, we classify Wannier localizability of bulk bands based on the homomorphisms from topological phases in the tenfold Altland-Zirnbauer (AZ) symmetry classes to those in the threefold Wigner-Dyson (WD) classes.
Notably, these two approaches from the boundary and bulk perspectives result in the same classification.
We clarify this agreement and develop a unified understanding of the bulk-boundary correspondence on the basis of $K$-theory.

This work is organized as follows.
In Sec.~\ref{sec: overview}, we provide an overview of this work before the detailed discussions.
In Sec.~\ref{sec: model}, we present a chiral-symmetric Hermitian lattice model with a flat and detached surface state in three dimensions.
In Sec.~\ref{sec: intrinsic}, we summarize intrinsic and extrinsic non-Hermitian topology, corresponding to nondetachable and detachable boundary states in Hermitian topological materials, respectively.
In Sec.~\ref{sec: WD}, we develop the classification of Wannier localizability based on the homomorphisms from the AZ classification to the WD classification.
Crucially, these two approaches from the boundary and bulk perspectives lead to the same classification (Table~\ref{tab: classification}).
In Sec.~\ref{sec: K-theory}, we clarify this relationship, using $K$-theory.
In Sec.~\ref{sec: conclusion}, we conclude this work.

%%%%% Overview %%%%%
\section{Overview}
    \label{sec: overview}

$K$-theory provides a powerful mathematical framework that effectively translates topological problems into algebraic ones, forming the foundation of the tenfold classification of topological insulators and superconductors~\cite{Kitaev-09, CTSR-review}.
Using $K$-theory, we clarify the relationship between the detachability of topological boundary states and Wannier localizability. 
In Sec.~\ref{sec: model}, we provide a prototypical chiral-symmetric model in three dimensions with a flat and detached surface state before the general classification.
For a gapped $d$-dimensional Hermitian system, we consider the bulk Hamiltonian $H(\bk)$ with the periodic boundary conditions and the boundary Hamiltonian $\hat H(\bk_\parallel)$ with the semi-infinite boundary conditions.
Here, $\bk \in T^d$ and $\bk_\parallel \in T^{d-1}$ are the $d$-dimensional wave vector in the bulk and $\left( d-1 \right)$-dimensional wave vector at the boundary, respectively, and $T^d$ and $T^{d-1}$ are the bulk and boundary Brillouin zone tori.
These bulk Hamiltonian $H(\bk)$ and boundary Hamiltonian $\hat H(\bk_\parallel)$ are respectively classified by the degree-$0$ $K$-group ${}^\phi K_G^{(\tau,c)+0}(T^d)$ and degree-($+1$) $K$-group ${}^\phi K_G^{(\tau,c)+1}(T^{d-1})$~\cite{Karoubi, AS-69, FM-13, Thiang-16, Shiozaki-17, Gomi-21}.
Here, $G$ is a symmetry group that is compatible with the boundary and possibly includes crystalline symmetry.
Moreover, $\phi: G\to \Z_2$ is a homomorphism specifying whether $g \in G$ is unitary symmetry or antiunitary symmetry, and $c: G\to \Z_2$ is a homomorphism specifying whether $g \in G$ is symmetry or antisymmetry (antisymmetry includes chiral symmetry and particle-hole symmetry).
Additionally, $\tau$ denotes a twist specifying the projective representation for internal degrees of freedom and nonprimitive lattice translations of space group symmetry.
Notably, while the Bloch Hamiltonian $H(\bk)$ is a finite-dimensional matrix, the boundary Hamiltonian $\hat H(\bk_\parallel)$ is an infinite-dimensional operator.
In terms of the $K$-groups, the bulk-boundary correspondence is denoted by 
\begin{equation}
\begin{array}{rccc}
f^{\rm BBC}_{G}: 
&{}^\phi K_G^{(\tau,c)+0}(T^d) &\longrightarrow& {}^\phi K_{G}^{(\tau,c)+1}(T^{d-1})                     \\
        & \rotatebox{90}{$\in$}&               & \rotatebox{90}{$\in$} \\
        & [H(\bk)]                    & \longmapsto   & [\hat H(\bk_\parallel)]
\end{array}
\end{equation}

We map the boundary Hermitian Hamiltonian $\hat H(\bk_\parallel)$ to a point-gapped non-Hermitian Hamiltonian $H_{P}$ by
\begin{equation}
H_{P} (\bk_\parallel) \coloneqq \ii \exp \left[ -\ii \pi \chi\,(\hat H(\bk_\parallel)) \right],
    \label{eq: HP}
\end{equation}
where $\chi(E)$ is a continuous function  defined as
\begin{equation}
    \chi(E) \coloneqq \begin{cases}
        {\rm sgn}(E) & |E| > E_{\rm gap}; \\
        E/E_{\rm gap} & |E| \leq E_{\rm gap}
    \end{cases}
        \label{eq: chi}
\end{equation}
with a bulk energy gap $E_{\rm gap} > 0$.
As shown in our accompanying Letter~\cite{NSSSK-24}, the detachment of $\hat H(\bk_\parallel)$ corresponds to imposing an imaginary line gap on $H_{P} (\bk_\parallel)$, further meaning that $H_{P}(\bk_\parallel)$ is continuously deformable to an anti-Hermitian Hamiltonian [i.e., $H_{P}^{\dag} (\bk_\parallel) = -H_{P} (\bk_\parallel)$]~\cite{KSUS-19}.
In Sec.~\ref{sec: intrinsic}, we review intrinsic (extrinsic) point-gap topology irreducible (reducible) to Hermitian or anti-Hermitian topology~\cite{OKSS-20, Shiozaki}, where we introduce a homomorphism $f^{\rm bdy}_{\rm L_i}$ that maps imaginary-line-gapped phases to point-gapped phases.
Notably, antisymmetries for $H_P(\bk_\parallel)$ behave as symmetries for $\ii H_P(\bk_\parallel)$, and eventually, the imaginary-line-gapped phases are classified by the degree-$0$ $K$-group with trivial $c$ [i.e., ${}^\phi K_{G}^{(\tau,c_{\rm triv})+0}(T^{d-1})$]. 
As demonstrated in Sec.~\ref{sec: K-theory}, if $G$ includes antisymmetry, we have the following commutative diagram: 
\begin{equation}
\begin{CD}
@.{}^\phi K_G^{(\tau,c_{\rm triv})+0}(T^{d-1}) \\
@. @VVf_{\rm L_i}^{\rm bdy}V \\
{}^\phi K_G^{(\tau,c)+0}(T^{d}) @>f_G^{\rm BBC}>> {}^\phi K_{G}^{(\tau,c)+1}(T^{d-1}) \\
@VVf^{\rm bulk}_0V @VVf^{\rm bdy}_0V \\
{}^\phi K_{G_0}^{\tau+0}(T^{d}) @>f_{G_0}^{\rm BBC}>> {}^\phi K_{G_0}^{\tau+1}(T^{d-1}) \\
\end{CD}
    \label{eq:exact_seq}
\end{equation}
Here, $G_0\subsetneq G$ is the subgroup excluding antisymmetries, the $K$-groups ${}^\phi K_{G_0}^{\tau+0}(T^{d})$ 
and ${}^\phi K_{G_0}^{\tau+1}(T^{d-1})$ respectively classify the bulk and boundary Hamiltonians protected only by $G_0$, and $f^{\rm bulk}_0$ and $f^{\rm bdy}_0$ are defined by forgetting antisymmetries. 
The right vertical line in Eq.~(\ref{eq:exact_seq}) is exact. 
Saliently, the exactness ${\rm im}\, f^{\rm bdy}_{\rm L_i} = {\rm ker}\, f^{\rm bdy}_0$ indicates that the boundary Hamiltonian $\hat H(\bk_\parallel)$ is detachable and exhibits no spectral flow ($[\hat H(\bk_\parallel)] \in {\rm im}\, f_{L_i}$) if and only if $\hat H(\bk_\parallel)$ is gappable ($[\hat H(\bk_\parallel)] \in {\rm ker}\, f_0$) when ignoring antisymmetries.

Suppose that the bulk Hamiltonian $H(\bk)$ exhibits spectral flow on the boundary, expressed as $f^{\rm BBC}_G([H(\bk)]) \notin {\rm im}\, f^{\rm bdy}_{L_{\rm i}}$. 
Then, the diagram in Eq.~(\ref{eq:exact_seq}) implies that $H(\bk)$ exhibits nontrivial topology in the $K$-group ${}^\phi K^{\tau+0}_{G_0}(T^d)$, 
which obstructs the existence of exponentially localized Wannier functions. 
This aligns with the fact that if a real-space boundary hosts a spectral flow, either occupied or unoccupied Bloch states are no longer Wannier localizable, as also argued in Ref.~\cite{Altland-24}.
Such a boundary includes cylindrical or spherical boundaries that can realize higher-order topological boundary states.
Furthermore, Table~\ref{tab: classification} is consistent with the symmetry-forgetting homomorphisms $f^{\rm bdy}_0$ for the tenfold AZ symmetry classes, as discussed in Sec.~\ref{sec: WD}.

%%%%%%%%%%%%%
\section{A model of detached and flat-band surface state}
    \label{sec: model}

We begin with a prototypical model exhibiting detachable topological boundary states. 
Specifically, we present a Hermitian lattice model with a flat and detached surface state for three dimensions $d=3$ in class AIII. 
The construction here can be straightforwardly generalized to any odd spatial dimension $d=2n+1$ with the even winding numbers $W_{2n+1} \in 2 \mathbb{Z}$.

%%%%%%%%%%%%%
\subsection{Model}

Consider the following four-band model in class AIII:
\begin{align}\label{eq:3DAIIImodel}
H(\bm{h}(k_x,k_y),k_z)
&= \left( \cos k_z \right) \tau_x \otimes \sigma_0 \nonumber \\
&\qquad + \left( \sin k_z \right) \tau_y \otimes \bm{h}(k_x,k_y)\cdot \bm{\sigma},
\end{align}
with 
\begin{equation}
    \bm{h}(k_x,k_y) \coloneqq (\sin k_x, \sin k_y, m - \cos k_x - \cos k_y).
\end{equation}
This lattice model indeed respects chiral symmetry
\begin{equation}
    \Gamma H(\bm{h}(k_x,k_y),k_z) \Gamma^{-1} = - H(\bm{h}(k_x,k_y),k_z),
\end{equation}
with $\Gamma = \tau_z$, and hence belongs to class AIII~\cite{CTSR-review}.
By writing the Hamiltonian as 
\begin{align}
H(\bm{h}(k_x,k_y),k_z) = \begin{pmatrix}
0&q^\dag(\bk) \\
q(\bk)&0 \\
\end{pmatrix}, 
\end{align}
with 
\begin{equation}
    q(\bk) \coloneqq \cos k_z + \ii \left( \sin k_z \right) \bm{h}(k_x,k_y)\cdot \bm{\sigma},
\end{equation}
the three-dimensional winding number is obtained as
\begin{align}
W_3 &= \frac{1}{24 \pi^2} \int_{T^3} \tr[q^{-1}dq]^3 \nonumber \\ 
&= \left\{
\begin{array}{ll}
0 & (|m| > 2); \\
-2 & (0 < m < 2); \\
2 & (-2 < m < 0). \\
\end{array}
\right.
\end{align}
This reflects twice the Chern number of the Hamiltonian $\bm{h}(\bk) \cdot \bm{\sigma}$ for two dimensions $d=2$ and class A.

\begin{figure*}[tbp]
    \centering
    \includegraphics[width=\linewidth]{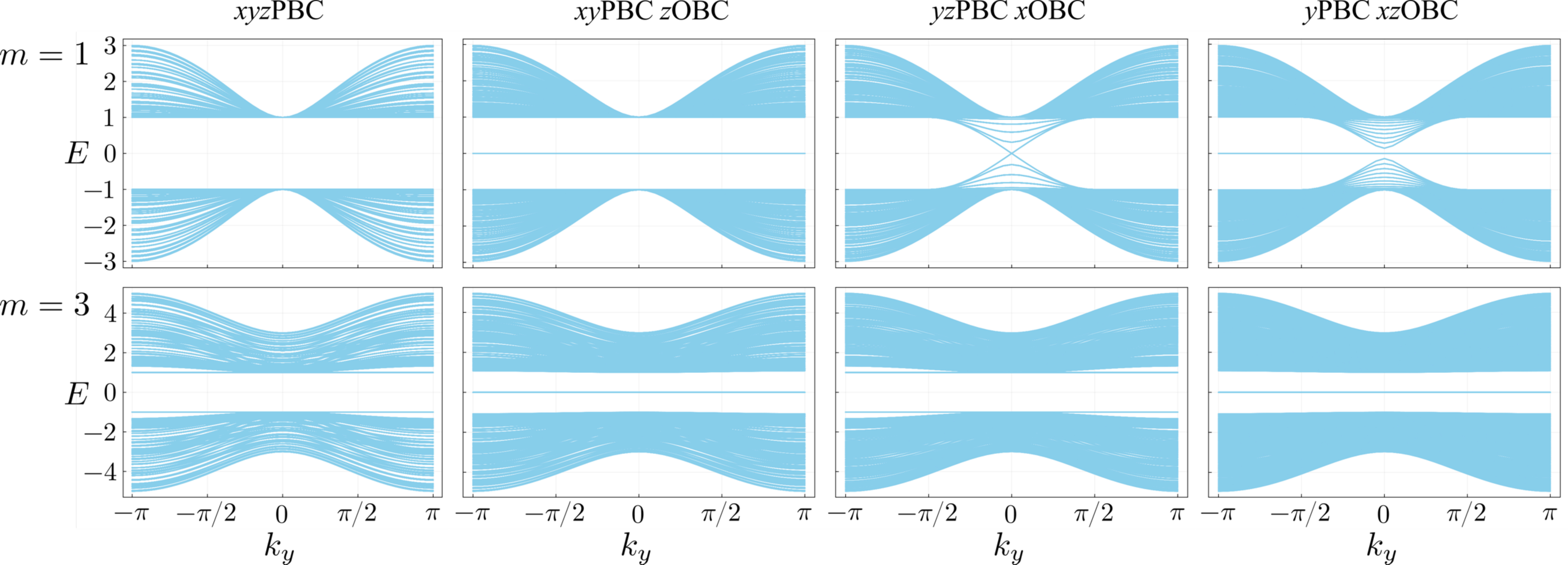} 
    \caption{Spectra of the model in Eq.~\eqref{eq:3DAIIImodel} for the topological phase ($m=1$; $W_3=-2$) and trivial phase ($m=3$; $W_3=0$).
    The periodic boundary conditions are imposed along the $y$ direction, by which the wave number $k_y$ is introduced.
    The other boundary conditions are (i)~the periodic boundary conditions along the $x$ and $z$ directions, (ii)~the periodic boundary conditions along the $x$ direction and open boundary conditions along the $z$ direction, (iii)~the periodic boundary conditions along the $z$ direction and open boundary conditions along the $x$ direction, and (iv)~the open boundary conditions along the $x$ and $z$ directions. 
    The system size along the $x$ and $z$ directions is taken as $L_x=L_z=21$. }
        \label{fig:surface_flat_band}
\end{figure*}

%%%%%%%%%%%%%
\subsection{Detached flat-band surface state}

Viewing $\bm{h}$ as a parameter, we consider the following one-dimensional lattice model along the $z$ direction:
\begin{align}
\hat{H}_{\rm 1D}(\bm{h}) = \sum_{z \in \mathbb{Z}} \psi^\dag_{z+1} \frac{\tau_x + \ii \tau_y \bm{h} \cdot \bm{\sigma}}{2} \psi_z + {\rm H.c.}, 
\end{align}
where $\psi_z$ is a four-component complex fermion operator at site $z$. 
Under the semi-infinite boundary conditions where $z \leq 0$ is in the topological insulator and $z > 0$ is in the vacuum, for every $|\bm{h}| > 0$, the interface $z=0$ hosts two exponentially localized zero-energy states, given as
\begin{align}
\phi_+(z, \bm{h}) &= u_+(\hat{h}) \otimes
\begin{pmatrix}
0 \\
\eta_-(z; h) \\
\end{pmatrix}_\tau, \\
\phi_-(z, \bm{h}) &= u_-(\hat{h}) \otimes 
\begin{pmatrix}
\eta_+(z; h) \\
0 \\
\end{pmatrix}_\tau,
\end{align}
with $h \coloneqq |\bm{h}|$ and $\hat{h} \coloneqq \bm{h} / |\bm{h}|$. 
Here, $u_{\pm}(\hat{h}) \in \mathbb{C}^2$ is a normalized two-component vector in the $\sigma$ space, satisfying 
\begin{equation}
\left( \bm{h} \cdot \bm{\sigma} \right) u_{\pm}(\hat{h}) = \pm h u_{\pm}(\hat{h}),
\end{equation}
and $(0~\eta_{-}(z; h))_\tau^T$ and $(\eta_{+}(z; h)~0)_\tau^T$ are respectively normalized vectors in the $\tau$ and $z$ spaces whose localization lengths depend on $h$. 
Note that $\phi_+(z, \bm{h})$ and $\phi_-(z, \bm{h})$ have chirality $\Gamma = 1$ and $\Gamma = -1$, respectively. 
With the polar coordinates of $\hat{h}$ as in $\hat{h} = (\theta_h, \phi_h)$, $u_\pm(\hat{h})$ is explicitly given, up to a $U(1)$ phase, as
\begin{align}
u_+(\hat{h}) 
&\propto
\begin{pmatrix}
\cos \left( \theta_h/2 \right) \\
e^{\ii \phi_h} \sin \left( \theta_h/2 \right)
\end{pmatrix}_\sigma, \\ 
u_-(\hat{h})
&\propto
\begin{pmatrix}
-e^{-\ii \phi_h} \sin \left( \theta_h/2 \right) \\
\cos \left( \theta_h/2 \right) \\
\end{pmatrix}_\sigma.
\end{align}
Thus, we obtain the two-component flat detached surface state, irrespective of the winding number $W_3$. 
However, its Chern number depends on that of $\bm{h}(\bk) \cdot \bm{\sigma}$, which further coincides with $W_3$.

We numerically calculate the spectrum of this model, as shown in Fig.~\ref{fig:surface_flat_band}.
Consistent with the above discussion, flat topological surface states detached from the bulk bands appear under the open boundary conditions along the $z$ direction.
By contrast, the topological surface states are connected to the bulk bands under the open boundary conditions along the $x$ direction.
It is also notable that flat surface states appear even in the topologically trivial phase (i.e., $m=3$; $W_3 = 0$).
However, these surface states do not exhibit the nontrivial Chern number.
Consequently, they are topologically unstable and can be removed from zero energy, as shown in the next subsection.

%%%%%%%%%%%%%
\subsection{Topological stability}

The bulk topology, three-dimensional winding number $W_3$, guarantees the stability of the Dirac surface states at zero energy.
Here, we demonstrate in a general manner that this topological stability persists even if the Dirac surface states are detached from the bulk bands.
Now, consider a generic perturbation to this flat-band surface state. 
The effective $2\times 2$ Hamiltonian with a chiral-symmetry-preserving perturbation $H_V(\bm{h})$ is
\begin{align}
H_{\rm eff}(\bm{h})= 
\int dz 
\begin{pmatrix}
    \phi_+^{\dag} (z, \bm{h}) \\ \phi_-^{\dag} (z, \bm{h})
\end{pmatrix}
H_V(\bm{h}) 
\begin{pmatrix}
    \phi_+(z, \bm{h}) & \phi_-(z, \bm{h})
\end{pmatrix}.
\end{align}
Notably, while $H_V(\bm{h})$ is gauge-independent, $\phi_{\pm}(z, \bm{h})$'s are gauge-dependent, and thus the effective Hamiltonian $H_{\rm eff}(\bm{h})$ is also gauge-dependent. 
Express $H_V(\bm{h})$ in the $\tau$ space as
\begin{align}
H_V(\bm{h}) = \begin{pmatrix}
0&V^\dag(\bm{h}) \\
V(\bm{h}) & 0 \\
\end{pmatrix}_\tau,
\end{align}
and introduce the $2 \times 2$ matrix
\begin{align}
v(\bm{h})
=\int dz~\eta_-(z; h)^* V(\bm{h}) \eta_+(z; h),
\end{align}
then the effective Hamiltonian becomes
\begin{align}
H_{\rm eff}(\bm{h})
=\begin{pmatrix}
0&u_+^\dag(\hat{h}) v^\dag(\bm{h}) u_-(\hat{h}) \\
u_-^\dag(\hat{h}) v(\bm{h}) u_+(\hat{h}) & 0 \\
\end{pmatrix}.
\end{align}
Thus, the spectrum of the surface states is 
\begin{align}
\pm |z(\bm{h})|,\quad 
z(\bm{h}) = u_-^\dag(\hat{h}) v(\bm{h}) u_+(\hat{h}) \in \mathbb{C}.
\end{align}
Note that the absolute value $|z(\bm{h})|$ is gauge-invariant.
Now, we show the following proposition:

\smallskip
{\it Proposition}.---There is no perturbation $v(\bm{h})$ such that $|z(\bm{h})| > 0$ for all $\hat{h} \in S^2$.

%\noindent
\smallskip
{\it Proof}.---We rewrite $|z|^2$ in the following manner, 
\begin{align}
|z|^2
&= \tr[v u_+ u_+^\dag v^\dag u_- u_-^\dag] \nonumber \\ 
&= \tr \left[ v \frac{1 + \hat{h} \cdot \bm{\sigma}}{2} v^\dag \frac{1 - \hat{h} \cdot \bm{\sigma}}{2} \right] \nonumber \\
&=\frac{1}{4} \tr [ vv^\dag + (v^\dag v - vv^\dag)\hat{h} \cdot \bm{\sigma} - v \hat{h} \cdot \bm{\sigma} v^\dag \hat{h} \cdot \bm{\sigma} ].
\end{align}
Using $v = v_0 + \bm{v} \cdot \bm{\sigma}$, we have 
\begin{align}
|z|^2 = (\hat{h} \times \bm{v}) \cdot (\hat{h} \times \bm{v}^*) - \ii \hat{h} \cdot (\bm{v}^* \times \bm{v})
\end{align}
from straightforward calculations.
We decompose $\bm{v}$ into the component of the polar coordinate $\bm{v} = v_h \hat{h} + v_\theta \hat{\theta} + v_\phi \hat{\phi}$ and obtain
\begin{align}
|z|^2 = |v_\theta - \ii v_\phi|^2.
\end{align}
Further separating $\bm{v}$ into real and imaginary parts $\bm{v} = \bm{v}_1 + \ii \bm{v}_2$ leads to
\begin{align}
|z|^2 = (v_{1\theta} + v_{2\phi})^2 + (v_{2\theta} - v_{1\phi})^2.
\end{align}
Introducing vectors on the tangent space $TS^2$,
\begin{align}
\bm{v}_1^\perp = \bm{v}_1 - (\hat{h} \cdot \bm{v}_1) \hat{h},\quad 
\bm{v}_2^\perp = \bm{v}_2 - (\hat{h} \cdot \bm{v}_2) \hat{h},
\end{align}
we eventually obtain the following form:
\begin{align}
|z|^2 = |\bm{v}_1^\perp - \hat{h} \times \bm{v}_2^\perp|^2.
\end{align}
The vector $\bm{v}_1^\perp - \hat{h} \times \bm{v}_2^\perp$ is in the tangent space $TS^2$. 
According to the Poincar\'e-Hopf theorem, there must be two defects on $S^2$ where the tangent vector $\bm{v}_1^\perp - \hat{h} \times \bm{v}_2^\perp$ vanishes. $\Box$

%\noindent
\smallskip
{\it Alternative proof}.---Note that $u_+(\hat{h})$ and $u_-(\hat{h})$ are orthonormal bases of $\mathbb{C}^2$. 
The $2 \times 2$ matrix $v(\bm{h})$ is expanded locally in $S^2$ as
\begin{align}
v(\bm{h}) &= \sum_{i, j \in \{+,-\}} v_{ij}(\bm{h}) u_i(\hat{h}) u_j^\dag(\hat{h}), \\
v_{ij}(\bm{h}) &\coloneqq u_i^\dag(\hat{h}) v(\bm{h}) u_j(\hat{h}),
\end{align}
so that $z(\bm{h}) = v_{-+}(\bm{h})$. 
Note that
\begin{align}
u_i(\hat{h}) u_j^\dag(\hat{h}) \in {\rm Mat}_{2 \times 2}(\mathbb{C}) \cong \mathbb{C}^2 \otimes (\mathbb{C}^2)^* 
\end{align}
defines a line bundle over $S^2$ with the fiber space ${\rm Mat}_{2 \times 2}(\mathbb{C})$. 
The line bundle defined by $u_-(\hat{h}) u_+^\dag(\hat{h})$ has the Chern number of 2. 
Assuming $v_{-+}(\bm{h}) \neq 0$ everywhere on $S^2$ means that there exists a global section of the nontrivial line bundle $u_-(\hat{h}) u_+^\dag(\hat{h})$, leading to a contradiction. $\Box$

A consequence of this statement is that if the detached surface state acquires a uniform mass gap in the whole surface Brillouin zone $(k_x, k_y) \in T^2$, the degree of the map $\hat{h}: T^2 \to S^2$, which coincides with the Chern number of the two-dimensional Hamiltonian $\bm{h}(k_x, k_y) \cdot \bm{\sigma}$ in class A and thus the winding number $W_3$ of the three-dimensional bulk, must be trivial.

\begin{table*}[t]
\caption{Homomorphisms $f_r, f_i$ from line-gap to point-gap topology for the tenfold Altland-Zirnbauer$^{\dag}$ classification.
Reproduced from Tables~S1 and S2 in the Supplemental Material of Ref.~\cite{OKSS-20}.}
\label{tab:SFH_AZdag}
\centering
{\scriptsize
$$
\begin{array}{ccccccccccccc}
\hline \hline
\mbox{Class} & \mbox{Gap} & d=0&d=1&d=2&d=3&d=4&d=5&d=6&d=7 \\
\hline
{\rm A}& {\rm L} \to {\rm P} &\Z \to 0&0 \to \Z&\Z \to 0&0 \to \Z&\Z \to 0&0 \to \Z&\Z \to 0&0 \to \Z\\ 
&&&&&&&&&\\ 
\hline
{\rm AIII}& {\rm L_r}\to {\rm P}&0 \to \Z&\Z \to 0&0 \to \Z&\Z \to 0&0 \to \Z&\Z \to 0&0 \to \Z&\Z \to 0\\
&&&&&&&&&\\ 
&{\rm L_i} \to {\rm P}&\Z\oplus\Z\to\Z&0\to 0&\Z\oplus\Z\to\Z&0\to 0&\Z\oplus\Z\to\Z&0\to 0&\Z\oplus\Z\to\Z&0\to 0\\ 
&&(n,m)\mapsto n-m&&(n,m)\mapsto n-m&&(n,m)\mapsto n-m&&(n,m)\mapsto n-m&\\ 
\hline
{\rm AI^\dag}& {\rm L} \to {\rm P}&\Z \to 0&0\to 0&0\to 0&0\to 2\Z&2\Z\to 0&0\to \Z_2&\Z_2\to \Z_2&\Z_2\to\Z\\ 
&&&&&&&&n\mapsto 0&\\
\hline
{\rm BDI^\dag}&{\rm L_r} \to {\rm P}&\Z_2 \to \Z&\Z \to 0&0\to 0&0\to 0&0\to 2\Z&2\Z\to 0&0\to \Z_2&\Z_2\to \Z_2\\ 
&&&&&&&&&n\mapsto 0\\ 
&{\rm L_i} \to {\rm P}&\Z\oplus \Z\to\Z&0\to 0&0\to 0&0\to 0&2\Z\oplus 2\Z\to 2\Z&0\to 0&\Z_2\oplus\Z_2\to\Z_2&\Z_2\oplus\Z_2\to\Z_2\\ 
&&(n,m)\mapsto n-m&&&&(n,m)\mapsto n-m&&(n,m)\mapsto n+m&(n,m)\mapsto n+m\\ 
\hline
{\rm D^\dag}&{\rm L_r} \to {\rm P}&\Z_2 \to \Z_2&\Z_2 \to \Z&\Z \to 0&0\to 0&0\to 0&0\to 2\Z&2\Z\to 0&0\to \Z_2\\ 
&&n \mapsto 0&&&&&&&\\
& {\rm L_i} \to {\rm P}&\Z \to \Z_2&0 \to \Z&0 \to 0&0 \to 0&2\Z\to 0&0\to 2\Z&\Z_2\to 0&\Z_2\to \Z_2 \\ 
&&n \mapsto n&&&&&&&n\mapsto n \\ 
\hline
{\rm DIII^\dag}& {\rm L_r} \to {\rm P}&0 \to \Z_2&\Z_2 \to \Z_2&\Z_2 \to \Z&\Z \to 0&0\to 0&0\to 0&0\to 2\Z&2\Z\to 0\\ 
&&&n \mapsto 0&&&&&&\\
&{\rm L_i} \to {\rm P}&\Z\to\Z_2&0\to\Z_2&\Z\to\Z&0\to 0&\Z\to 0&0\to 0&\Z\to 2\Z&0\to 0\\ 
&&n\mapsto n&&n\mapsto 2n&&&&n\mapsto n&\\ 
\hline
{\rm AII^\dag}& {\rm L} \to {\rm P}&2\Z \to 0&0 \to \Z_2&\Z_2 \to \Z_2&\Z_2 \to \Z&\Z \to 0&0\to 0&0\to 0&0\to 2\Z\\
&&&&n \mapsto 0&&&&&\\
\hline
{\rm CII^\dag}& {\rm L_r} \to {\rm P}&0 \to 2\Z&2\Z \to 0&0 \to \Z_2&\Z_2 \to \Z_2&\Z_2 \to \Z&\Z \to 0&0\to 0&0\to 0\\ 
&&&&&n \mapsto 0&&&&\\
&{\rm L_i} \to {\rm P}&2\Z\oplus 2\Z\to 2\Z&0\to 0&\Z_2\oplus\Z_2\to \Z_2&\Z_2\oplus \Z_2\to\Z_2&\Z\oplus \Z\to\Z&0\to 0&0\to 0&0\to 0\\ 
&&(n,m)\mapsto n-m&&(n,m)\mapsto n+m&(n,m)\mapsto n+m&(n,m)\mapsto n-m&&\\
\hline
{\rm C^\dag}& {\rm L_r} \to {\rm P}&0 \to 0&0 \to 2\Z&2\Z \to 0&0 \to \Z_2&\Z_2 \to \Z_2&\Z_2 \to \Z&\Z \to 0&0\to 0\\ 
&&&&&&n \mapsto 0&&&\\ 
& {\rm L_i} \to {\rm P}&2\Z \to 0&0\to 2\Z&\Z_2 \to 0&\Z_2 \to \Z_2&\Z \to \Z_2&0 \to \Z&0 \to 0&0 \to 0\\ 
&&&&&n \mapsto n&n \mapsto n&&&\\ 
\hline
{\rm CI^\dag}& {\rm L_r} \to {\rm P}&0 \to 0&0 \to 0&0 \to 2\Z&2\Z \to 0&0 \to \Z_2&\Z_2 \to \Z_2&\Z_2 \to \Z&\Z \to 0\\ 
&&&&&&&n \mapsto 0&&\\
& {\rm L_i} \to {\rm P}&\Z\to 0&0\to 0&\Z\to 2\Z&0\to 0&\Z\to\Z_2&0\to\Z_2&\Z\to\Z&0\to 0\\ 
&&&&n\mapsto n&&n\mapsto n&&n\mapsto 2n&\\
\hline \hline
\end{array}
$$
}
\end{table*}

%%%%%%%%%%
\section{Intrinsic and extrinsic non-Hermitian topology}
    \label{sec: intrinsic}

As also discussed above, while intrinsic non-Hermitian topology prohibits the detachment of topological boundary states, extrinsic non-Hermitian topology permits it and further leads to anti-Hermitian topology of the detached boundary states~\cite{NSSSK-24}.
In this section, we discuss the classification of intrinsic and extrinsic non-Hermitian topology, as shown in Table~\ref{tab:SFH_AZdag} (see also Sec.~SIX in the Supplemental Material of Ref.~\cite{OKSS-20}, as well as Ref.~\cite{Shiozaki}).

%%%%%%%%%%
\subsection{General formalism}

In the presence of a point (line) gap, complex-energy bands are defined not to cross a reference point (line) in the complex-energy plane~\cite{KSUS-19}.
Even if a point gap is open, a line gap is not necessarily open.
However, a point gap is always open when a line gap is open by placing a reference point on the reference line.
Thus, we can introduce a map from line-gapped topological phases to point-gapped topological phases for each spatial dimension and symmetry class.
Given a $d$-dimensional non-Hermitian Hamiltonian $H \left( \bm{k} \right)$, we consider the Hermitized Hamiltonian
\begin{equation}
    \tilde{H} \left( \bm{k} \right) \coloneqq \begin{pmatrix}
        0 & H \left( \bm{k} \right) \\
        H^{\dag} \left( \bm{k} \right) & 0
    \end{pmatrix}_{\sigma},
        \label{eq: Hermitization}
\end{equation}
which respects chiral symmetry 
\begin{equation}
\Gamma H \left( \bm{k} \right) \Gamma^{-1} = - H \left( \bm{k} \right)
\end{equation}
with $\Gamma = \sigma_z$ by construction.
When the non-Hermitian Hamiltonian $H \left( \bm{k} \right)$ is point-gapped, the Hermitized Hamiltonian $\tilde{H} \left( \bm{k} \right)$ is also gapped, and vice versa.
Thus, the topological classification of $H \left( \bm{k} \right)$ coincides with that of $\tilde{H} \left( \bm{k} \right)$, which is denoted by $K_{\rm P}$.
By contrast, the presence of a line gap imposes an additional constraint on $\tilde{H} \left( \bm{k} \right)$.
In fact, a non-Hermitian Hamiltonian $H \left( \bm{k} \right)$ with a real or an imaginary line gap is continuously deformable to a Hermitian or an anti-Hermitian Hamiltonian, respectively, while keeping the line gap and relevant symmetry~\cite{KSUS-19}.
Notably, Hermiticity or anti-Hermiticity of $H \left( \bm{k} \right)$ imposes another chiral symmetry with $\Gamma_{\rm r} = \sigma_y$ or $\Gamma_{\rm i} = \sigma_x$ on the Hermitized Hamiltonian $\tilde{H} \left( \bm{k} \right)$, respectively.
This additional chiral symmetry leads to the different topological classification described by $K_{\rm L_{r}}$ or $K_{\rm L_{i}}$ in the presence of a real or an imaginary line gap (and a point gap as well).
Forgetting $\Gamma_{\rm r}$ ($\Gamma_{\rm i}$) defines a homomorphism $f_{\rm r}: K_{\rm L_r} \to K_{\rm P}$ ($f_{\rm i}: K_{\rm L_i} \to K_{\rm P}$) from $K_{\rm L_r}$ ($K_{\rm L_i}$) to $K_{\rm P}$.
Owing to the dimensional isomorphism of $K$-theory~\cite{Karoubi}, it is sufficient to compute $f_{\rm r}$ or $f_{\rm i}$ in zero dimension to obtain those in arbitrary dimensions.
Table~\ref{tab:SFH_AZdag} summarizes these homomorphisms in the AZ$^{\dag}$ symmetry classes, which is relevant to boundary states in Hermitian topological insulators and superconductors.
The homomorphisms in all the 38-fold symmetry classes are found in Refs.~\cite{OKSS-20, Shiozaki}.

If a point-gapped non-Hermitian Hamiltonian $H \left( \bm{k} \right)$ is included in the image of either homomorphism $f_{\rm r}$ or $f_{\rm i}$, it is continuously deformable to a Hermitian or an anti-Hermitian Hamiltonian, implying that the topological nature is also attributed to the conventional Hermitian one.
Conversely, if $H \left( \bm{k} \right)$ is not included in the image of $f_{\rm r}$ or $f_{\rm i}$, its topological nature is intrinsic to non-Hermitian systems.
Such intrinsic non-Hermitian topology is captured by the quotient group $K_{\rm P}/\left( \mathrm{Im}\,f_{\rm r} \cup \mathrm{Im}\,f_{\rm i}\right)$, which is also summarized in Table~\ref{tab:SFH_AZdag}.

%%%%%%%%%%
\subsection{Examples}

As a prime example of intrinsic non-Hermitian topology, we study class A, i.e., non-Hermitian systems without any internal symmetry.
In class A, while line-gap topology (or equivalently, Hermitian topology) is nontrivial only in even spatial dimensions, point-gap topology is nontrivial only in odd spatial dimensions~\cite{KSUS-19}.
Consequently, point-gap topology in class A is always intrinsic to non-Hermitian systems.
This intrinsic point-gap topology is the origin of the non-Hermitian skin effect~\cite{Lee-16, YW-18-SSH, Kunst-18, Yokomizo-19, Zhang-20, OKSS-20} and exceptional points~\cite{Shen-18, KBS-19, Denner-21, Nakamura-24}.
As discussed in Ref.~\cite{NSSSK-24}, it also underlies the spectral flow and nondetachability of chiral edge states in Chern insulators.

On the other hand, as an exemplary class of extrinsic non-Hermitian topology, let us study class AIII, i.e., non-Hermitian systems that respect chiral symmetry
\begin{equation}
    \Gamma H^{\dag} \Gamma^{-1} = - H.
        \label{eq: NH CS}
\end{equation}
Here, $\Gamma$ is a unitary matrix $\Gamma$ satisfying $\Gamma \Gamma^{\dag} = \Gamma^{\dag} \Gamma = \Gamma^2 = 1$.
In class AIII, point-gap topology is nontrivial and leads to the $\mathbb{Z}$ classification in even spatial dimensions;
real-line-gap topology results in the $\mathbb{Z}$ classification in odd spatial dimensions, whereas imaginary-line-gap topology the $\mathbb{Z} \oplus \mathbb{Z}$ classification in even spatial dimensions.
Consequently, the $\mathbb{Z}$ point-gap topology can be incorporated in the $\mathbb{Z} \oplus \mathbb{Z}$ imaginary-line-gap topology.

To clarify the relationship between point-gap and imaginary-line-gap topology, we investigate flattened chiral-symmetric non-Hermitian systems in zero dimension.
In the presence of an imaginary line gap, non-Hermitian systems can be flattened into anti-Hermitian systems (i.e., $H^{\dag} = - H$) while preserving chiral symmetry and an imaginary line gap.
Then, chiral symmetry behaves as $\mathbb{Z}_2$ unitary symmetry for anti-Hermitian systems $H$ (or equivalently, $\mathbb{Z}_2$ unitary symmetry for Hermitian systems $\ii H$).
As a result, the $K$-group $K_{\rm L_i}$ for the $\mathbb{Z} \oplus \mathbb{Z}$ topological classification is generated by 
\begin{align}
    \left( 1, 0 \right) \in \mathbb{Z} \oplus \mathbb{Z}: \quad \left[ \Gamma = +1, H_0 = \ii, H_1 = -\ii \right], \label{eq: AIII Z+Z 10} \\
    \left( 0, 1 \right) \in \mathbb{Z} \oplus \mathbb{Z}: \quad \left[ \Gamma = -1, H_0 = \ii, H_1 = -\ii \right], \label{eq: AIII Z+Z 01}
\end{align}
where $H_0$ and $H_1$ are topologically nontrivial and trivial flattened systems in terms of an imaginary line gap. 
In the presence of a point gap, by contrast, topology is captured by Hermitian matrices $\ii H \Gamma$ (see, for example, Sec.~SVII of the Supplemental Material in Ref.~\cite{KBS-19}).
Thus, the $K$-group $K_{\rm P}$ for the $\mathbb{Z}$ topological classification is generated by
\begin{equation}
    1 \in \mathbb{Z}: \quad \left[ \ii \Gamma H = -1, \ii \Gamma H = +1 \right].
        \label{eq: AIII Z 1}
\end{equation}
Comparing the $\mathbb{Z} \oplus \mathbb{Z}$ classification in Eqs.~(\ref{eq: AIII Z+Z 10}) and (\ref{eq: AIII Z+Z 01}) with the $\mathbb{Z}$ classification in Eq.~(\ref{eq: AIII Z 1}), we have a homomorphism
\begin{equation}
\begin{array}{rccc}
f_{\rm i}: 
&\mathbb{Z} \oplus \mathbb{Z}~\left( K_{\rm L_i} \right) &\longrightarrow& \mathbb{Z}~\left( K_{\rm P} \right)                   \\
        & \rotatebox{90}{$\in$}&               & \rotatebox{90}{$\in$} \\
        & \left( n, m \right)                    & \longmapsto   & n-m
\end{array}
\end{equation}
Therefore, the $\mathbb{Z}$ point-gap topology is continuously deformable to a part of the $\mathbb{Z} \oplus \mathbb{Z}$ imaginary-line-gap topology.
This extrinsic nature of point-gap topology leads to the detachment of topological boundary states in class AIII, as discussed in Ref.~\cite{NSSSK-24}.

Notably, the zero-dimensional analysis is generally extended to arbitrary spatial dimensions, ensured by the dimensional isomorphism of $K$-theory~\cite{Karoubi}.
In a similar manner, the homomorphisms from line-gap topology to point-gap topology are identified in the tenfold AZ$^{\dag}$ symmetry classes, as summarized in Table~\ref{tab:SFH_AZdag}.

\begin{table*}[tbp]
	\centering
    %\hspace*{-1em}
	\caption{Topological invariants for the tenfold Altland-Zirnbauer classes. There exist four types of invariants: For the $\mathbb{Z}$ indices, ``$W_d$" is the $d$-dimensional winding number and ``$\mathrm{Ch}_{d/2}$" is the $d/2$-th Chern number, which are defined in odd and even $d$ spatial dimensions, respectively.
    For the $\mathbb{Z}_2$ indices, ``$\mathrm{FK}_{d/2}$" denotes the Fu-Kane invariant in even $d$ dimensions, and ``$\mathrm{CS}_{d}~(\widetilde{\mathrm{CS}}_d)$" represents the Chern-Simons invariant in symmetry classes without (with) chiral symmetry in odd $d$ dimensions. 
    The entries ``$0$" indicate the absence of nontrivial topological phases.}
	\label{atab:AZ invariants}
 {\scriptsize
     \begin{tabular}{c|ccc|cccccccc} \hline \hline
    ~~Class~~ & ~~TRS~~ & ~~PHS~~ & ~~CS~~ & ~~$d=1$~~ & ~~$d=2$~~ & ~~$d=3$~~ & ~~$d=4$~~ & ~~$d=5$~~ & ~~$d=6$~~ & ~~$d=7$~~ & ~~$d=8$~~ \\ \hline
    A & $0$ & $0$ & $0$ & $0$ & ~~$\mathrm{Ch}_1\in\mathbb{Z}^{\checkmark}$~~ & $0$ & ~~$\mathrm{Ch}_2\in\mathbb{Z}^{\checkmark}$~~ & $0$ & ~~$\mathrm{Ch}_3\in\mathbb{Z}^{\checkmark}$~~ & $0$ & ~~$\mathrm{Ch}_4\in\mathbb{Z}^{\checkmark}$~~ \\
    AIII & $0$ & $0$ & $1$ & ~~$W_1\in\mathbb{Z}^{\times}$~~ & $0$ & ~~$W_3\in\mathbb{Z}^{\times}$~~ & $0$ & ~~$W_5\in\mathbb{Z}^{\times}$~~ & $0$ & ~~$W_7\in\mathbb{Z}^{\times}$~~ & $0$ \\ \hline
    AI & $+1$ & $0$ & $0$ & $0$ & $0$ & $0$ & $\mathrm{Ch}_2\in2\mathbb{Z}^{\checkmark}$ & $0$ & $\mathrm{FK}_3\in\mathbb{Z}_2^{\checkmark}$ & $\mathrm{CS}_7\in\mathbb{Z}_2^{\checkmark}$ & $\mathrm{Ch}_4\in\mathbb{Z}^{\checkmark}$ \\
    BDI & $+1$ & $+1$ & $1$ & $W_1\in\mathbb{Z}^{\times}$ & $0$ & $0$ & $0$ & $W_5\in2\mathbb{Z}^{\times}$ & $0$ & $\widetilde{\mathrm{CS}}_7\in\mathbb{Z}_2^{\times}$ & $\mathrm{FK}_4\in\mathbb{Z}_2^{\times}$ \\
    D & $0$ & $+1$ & $0$ & $\mathrm{CS}_1\in\mathbb{Z}_2^{\times}$ & $\mathrm{Ch}_1\in\mathbb{Z}^{\checkmark}$ & $0$ & $0$ & $0$ & $\mathrm{Ch}_3\in2\mathbb{Z}^{\checkmark}$ & $0$ & $\mathrm{FK}_4\in\mathbb{Z}_2^{\times}$ \\
    DIII & $-1$ & $+1$ & $1$ & $\widetilde{\mathrm{CS}}_1\in\mathbb{Z}_2^{\times}$ & $\mathrm{FK}_1\in\mathbb{Z}_2^{\checkmark}$ & $W_3\in\mathbb{Z}^{\checkmark/\times}$ & $0$ & $0$ & $0$ & $W_7\in2\mathbb{Z}^{\times}$ & $0$ \\
    AII & $-1$ & $0$ & $0$ & $0$ & $\mathrm{FK}_1\in\mathbb{Z}_2^{\checkmark}$ & $\mathrm{CS}_3\in\mathbb{Z}_2^{\checkmark}$ & $\mathrm{Ch}_2\in\mathbb{Z}^{\checkmark}$ & $0$ & $0$ & $0$ & $\mathrm{Ch}_4\in2\mathbb{Z}^{\checkmark}$ \\
    CII & $-1$ & $-1$ & $1$ & $W_1\in2\mathbb{Z}^{\times}$ & $0$ & $\widetilde{\mathrm{CS}}_3\in\mathbb{Z}_2^{\times}$ & $\mathrm{FK}_2\in\mathbb{Z}_2^{\times} $ & $W_5\in\mathbb{Z}^{\times}$ & $0$ & $0$ & $0$ \\
    C & $0$ & $-1$ & $0$ & $0$ & $\mathrm{Ch}_1\in2\mathbb{Z}^{\checkmark}$ & $0$ & $\mathrm{FK}_2\in\mathbb{Z}_2^{\times}$ & $\mathrm{CS}_5\in\mathbb{Z}_2^{\times}$ & $\mathrm{Ch}_3\in\mathbb{Z}^{\checkmark}$ & $0$ & $0$ \\
    CI & $+1$ & $-1$ & $1$ & $0$ & $0$ & $W_3\in2\mathbb{Z}^{\times}$ & $0$ & $\widetilde{\mathrm{CS}}_5\in\mathbb{Z}_2^{\times}$ & $\mathrm{FK}_3\in\mathbb{Z}_2^{\checkmark}$ & $W_7\in\mathbb{Z}^{\checkmark/\times}$ & $0$ \\ \hline \hline
  \end{tabular}
  }
\end{table*}

\begin{table*}[tbp]
	\centering
    %\hspace*{-1em}
	\caption{Homomorphisms from the tenfold Altland-Zirnbauer (AZ) symmetry classification to the threefold Wigner-Dyson (WD) classification.
    While the tenfold AZ classes are specified by time-reversal symmetry (TRS), particle-hole symmetry (PHS), and chiral symmetry (CS), the threefold WD classes are based solely on TRS.
    For the entries ``$\mathbb{Z}^{\checkmark} \xrightarrow{\cong} \mathbb{Z}$", ``$2\mathbb{Z}^{\checkmark} \xrightarrow{\cong} 2\mathbb{Z}$", and ``$\mathbb{Z}_2^{\checkmark}\xrightarrow{\cong}\mathbb{Z}_2$", the original topological phases in the AZ classes are isomorphic to those in the corresponding WD classes.
    For the entries ``$\mathbb{Z}_2^{\times} \xrightarrow{0} \mathbb{Z}_2$" and ``$\mathbb{Z}_2^{\times} \xrightarrow{0} \mathbb{Z}$", the original topological phases in the AZ classes vanish in the corresponding WD classes.
    For the entries ``$2\mathbb{Z}^{\checkmark} \xrightarrow{2} \mathbb{Z}$", the topological invariants double. 
    For the entries $\mathbb{Z}^{\checkmark/\times} \xrightarrow{\bmod 2} \mathbb{Z}_2$, the topological phases with the odd (even) number of topological invariants survive (vanish).
    }
	\label{atab: Hom}
 {\scriptsize
     \begin{tabular}{c|ccc|cccccccc} \hline \hline
    ~~Class~~ & ~~TRS~~ & ~~PHS~~ & ~~CS~~ & ~~$d=1$~~ & ~~$d=2$~~ & ~~$d=3$~~ & ~~$d=4$~~ & ~~$d=5$~~ & ~~$d=6$~~ & ~~$d=7$~~ & ~~$d=8$~~ \\ \hline
    A $\rightarrow$ A & $0$ & $0$ & $0$ & $0 \rightarrow 0$ & ~~$\mathbb{Z}^{\checkmark} \xrightarrow{\cong} \mathbb{Z}$~~ & $0 \rightarrow 0$ & ~~$\mathbb{Z}^{\checkmark} \xrightarrow{\cong} \mathbb{Z}$~~ & $0 \rightarrow 0$ & ~~$\mathbb{Z}^{\checkmark} \xrightarrow{\cong} \mathbb{Z}$~~ & $0 \rightarrow 0$ & ~~$\mathbb{Z}^{\checkmark} \xrightarrow{\cong} \mathbb{Z}$~~ \\
    AIII $\rightarrow$ A & $0$ & $0$ & $1$ & ~~$\mathbb{Z}^{\times} \rightarrow 0$~~ & $0 \rightarrow \mathbb{Z}$ & ~~$\mathbb{Z}^{\times} \rightarrow 0$~~ & $0 \rightarrow \mathbb{Z}$ & ~~$\mathbb{Z}^{\times} \rightarrow 0$~~ & $0 \rightarrow \mathbb{Z}$ & ~~$\mathbb{Z}^{\times} \rightarrow 0$~~ & $0 \rightarrow \mathbb{Z}$ \\ \hline
    AI $\rightarrow$ AI & $+1$ & $0$ & $0$ & $0 \rightarrow 0$ & $0 \rightarrow 0$ & $0 \rightarrow 0$ & $2\mathbb{Z}^{\checkmark} \xrightarrow{\cong} 2\mathbb{Z}$ & $0 \rightarrow 0$ & $\mathbb{Z}_2^{\checkmark} \xrightarrow{\cong} \mathbb{Z}_2$ & $\mathbb{Z}_2^{\checkmark} \xrightarrow{\cong} \mathbb{Z}_2$ & $\mathbb{Z}^{\checkmark} \xrightarrow{\cong} \mathbb{Z}$ \\
    BDI $\rightarrow$ AI & $+1$ & $+1$ & $1$ & $\mathbb{Z}^{\times} \rightarrow 0$ & $0 \rightarrow 0$ & $0 \rightarrow 0$ & $0 \rightarrow 2\mathbb{Z}$ & $2\mathbb{Z}^{\times} \rightarrow 0$ & $0 \rightarrow \mathbb{Z}_2$ & $\mathbb{Z}_2^{\times} \xrightarrow{0} \mathbb{Z}_2$ & $\mathbb{Z}_2^{\times} \xrightarrow{0} \mathbb{Z}$ \\
    D $\rightarrow$ A & $0$ & $+1$ & $0$ & $\mathbb{Z}_2^{\times} \rightarrow 0$ & $\mathbb{Z}^{\checkmark} \xrightarrow{\cong} \mathbb{Z}$ & $0 \rightarrow 0$ & $0 \rightarrow \mathbb{Z}$ & $0 \rightarrow 0$ & $2\mathbb{Z}^{\checkmark} \xrightarrow{2} \mathbb{Z}$ & $0 \rightarrow 0$ & $\mathbb{Z}_2^{\times} \xrightarrow{0} \mathbb{Z}$ \\
    ~~DIII $\rightarrow$ AII~~ & $-1$ & $+1$ & $1$ & $\mathbb{Z}_2^{\times} \rightarrow 0$ & $\mathbb{Z}_2^{\checkmark} \xrightarrow{\cong} \mathbb{Z}_2$ & $\mathbb{Z}^{\checkmark/\times} \xrightarrow{\bmod 2} \mathbb{Z}_2$ & $0 \rightarrow \mathbb{Z}$ & $0 \rightarrow 0$ & $0 \rightarrow 0$ & $2\mathbb{Z}^{\times} \rightarrow 0$ & $0 \rightarrow 2\mathbb{Z}$ \\
    AII $\rightarrow$ AII & $-1$ & $0$ & $0$ & $0\to 0$ & $\mathbb{Z}_2^{\checkmark} \xrightarrow{\cong} \mathbb{Z}_2$ & $\mathbb{Z}_2^{\checkmark} \xrightarrow{\cong} \mathbb{Z}_2$ & $\mathbb{Z}^{\checkmark} \xrightarrow{\cong} \mathbb{Z}$ & $0\to 0$ & $0\to 0$ & $0\to 0$ & $2\mathbb{Z}^{\checkmark} \xrightarrow{\cong} 2\mathbb{Z}$ \\
    CII $\rightarrow$ AII & $-1$ & $-1$ & $1$ & $2\mathbb{Z}^{\times} \to 0$ & $0\to \mathbb{Z}_2$ & $\mathbb{Z}_2^{\times} \xrightarrow{0} \mathbb{Z}_2$ & $\mathbb{Z}_2^{\times} \xrightarrow{0} \mathbb{Z}$ & $\mathbb{Z}^{\times} \to 0$ & $0\to 0$ & $0\to 0$ & $0\to 2\mathbb{Z}$ \\
    C $\rightarrow$ A & $0$ & $-1$ & $0$ & $0 \to 0$ & $2\mathbb{Z}^{\checkmark} \xrightarrow{2} \mathbb{Z}$ & $0\to 0$ & $\mathbb{Z}_2^{\times} \xrightarrow{0} \mathbb{Z}$ & $\mathbb{Z}_2^{\times} \to 0$ & $\mathbb{Z}^{\checkmark} \xrightarrow{\cong} \mathbb{Z}$ & $0\to 0$ & $0\to \mathbb{Z}$ \\
    CI $\rightarrow$ AI & $+1$ & $-1$ & $1$ & $0 \to 0$ & $0\to 0$ & $2\mathbb{Z}^{\times} \to 0$ & $0\to 2\mathbb{Z}$ & $\mathbb{Z}_2^{\times} \to 0 $ & $\mathbb{Z}_2^{\checkmark}\xrightarrow{\cong}\mathbb{Z}_2$ & $\mathbb{Z}^{\checkmark/\times} \xrightarrow{\bmod 2} \mathbb{Z}_2$ & $0\to \mathbb{Z}$ \\ \hline \hline
  \end{tabular}
  }
\end{table*}

%%%%%%%%%%
\section{From Altland-Zirnbauer classification to Wigner-Dyson classification}
    \label{sec: WD}

While topology protected by no internal symmetry or time-reversal symmetry, including the Chern number, should impose Wannier obstructions, topology protected by chiral or particle-hole symmetry should be exempt from such Wannier obstructions, as argued in Sec.~II\,C of Ref.~\cite{Altland-24}.
Consequently, studying how essential chiral or particle-hole symmetry for given topological invariants, we identify the presence or absence of the Wannier obstructions, leading to their tenfold classification summarized in Table~\ref{tab: classification}.
To clarify it, we explicitly provide homomorphisms from topological phases in the tenfold AZ symmetry classes to those in the threefold WD classes (i.e., classes A, AI, and AII). 
We note that the classification of Wannier localizability was provided for one, two, and three dimensions also in Ref.~\cite{Altland-24}.

Let us first introduce the topological invariants for the AZ classification. 
As summarized in Table~\ref{atab:AZ invariants}, each nontrivial $d$-dimensional topological phase is characterized by one of $\{W_d, \mathrm{Ch}_{d/2}, \mathrm{FK}_{d/2}, \mathrm{CS}_d~ (\widetilde{\mathrm{CS}}_d)\}$, 
which are the $d$-dimensional winding number, the $d/2$-th Chern number, the Fu-Kane invariant, and the Chern-Simons invariant in symmetry classes without (with) chiral symmetry, respectively~\cite{CTSR-review}. 
The homomorphisms are given in Table~\ref{atab: Hom}, based on the connections among these topological invariants, i.e., how the $d$-dimensional topological invariants in the AZ classes relate to those in the WD classes if one virtually forgets chiral symmetry and/or particle-hole symmetry. 
To derive these homomorphisms, we divide all the entries in Table~\ref{atab: Hom} into the following seven categories. 
We identify the homomorphism for each case below.

%\begin{enumerate}
%\setcounter{enumi}{0}
%\item 

%%%%%%%%%%
\subsection{$\mathbb{Z}^{\checkmark} \xrightarrow{\cong} \mathbb{Z}$,~~ $2\mathbb{Z}^{\checkmark} \xrightarrow{\cong} 2\mathbb{Z}$,~~ $\mathbb{Z}_2^{\checkmark}\xrightarrow{\cong}\mathbb{Z}_2$}
%$\{$``$\mathbb{Z}^{\checkmark} \xrightarrow{\cong} \mathbb{Z}$", ``$2\mathbb{Z}^{\checkmark} \xrightarrow{\cong} 2\mathbb{Z}$", ``$\mathbb{Z}_2^{\checkmark}\xrightarrow{\cong}\mathbb{Z}_2$"$\}$

Let us focus on class DIII in two dimensions to explain how to see and derive Table~\ref{atab: Hom}. 
Since class DIII respects spinful time-reversal symmetry, it reduces to class AII if one ignores both chiral symmetry and particle-hole symmetry. 
See the first four columns of Table~\ref{atab:AZ invariants} and Table~\ref{atab: Hom}. 
The entries ``$\mathbb{Z}_2^{\checkmark}\xrightarrow{\cong}\mathbb{Z}_2$" show how the topological phase in class DIII, the left side of the arrow (i.e., ``$\mathbb{Z}_2^{\checkmark}$"), is mapped to that in class AII, the right side of the arrow (i.e., ``$\mathbb{Z}_2$"). 
Since, for two dimensions, both the $\mathbb{Z}_2$ invariants coincide and are given by the Fu-Kane invariant $\mathrm{FK}_1$ (see Table~\ref{atab:AZ invariants})~\cite{Fu-Kane-06}, we conclude that the nontrivial topological phase in two-dimensional class DIII is isomorphically mapped to that in two-dimensional class AII. 
We use the symbol ``$\cong$" in ``$\mathbb{Z}_2^{\checkmark}\xrightarrow{\cong}\mathbb{Z}_2$" to emphasize that this homomorphism is indeed an (group) isomorphism. 
In a similar manner, we 
have an isomorphism for each of ``$\mathbb{Z}^{\checkmark} \xrightarrow{\cong} \mathbb{Z}$", ``$2\mathbb{Z}^{\checkmark} \xrightarrow{\cong} 2\mathbb{Z}$", and  ``$\mathbb{Z}_2^{\checkmark}\xrightarrow{\cong}\mathbb{Z}_2$" in Table~\ref{atab: Hom}. 

%%%%%%%%%%
\subsection{$2\mathbb{Z}^{\checkmark} \xrightarrow{2} \mathbb{Z}$}
%\item $\{$``$2\mathbb{Z}^{\checkmark} \xrightarrow{2} \mathbb{Z}$"$\}$

The entries ``$2\mathbb{Z}^{\checkmark} \xrightarrow{2} \mathbb{Z}$" appear in two-dimensional class C and six-dimensional class D. 
These two classes reduce to class A when one forgets particle-hole symmetry. In two dimensions, the first Chern number characterizes both classes A and C, where the presence of particle-hole symmetry makes it even integers. 
Thus, we have a map from the topological phases in two-dimensional class C to the even parts of those in two-dimensional class A. 
This homomorphism is denoted by ``$2\mathbb{Z}^{\checkmark} \xrightarrow{2} \mathbb{Z}$". 
Similarly, one can show this kind of map for six-dimensional class D.

%%%%%%%%%%
\subsection{$\mathbb{Z}^{\checkmark/\times} \xrightarrow{\bmod 2} \mathbb{Z}_2$}
%\item $\{$``$\mathbb{Z}^{\checkmark/\times} \xrightarrow{\bmod 2} \mathbb{Z}_2$"$\}$

In three dimensions, we have the entry ``$\mathbb{Z}^{\checkmark/\times} \xrightarrow{\bmod 2} \mathbb{Z}_2$" for class DIII. 
Since the corresponding WD symmetry class is class AII, what we should know is the connection between the three-dimensional winding number $W_3$ and the Chern-Simons invariant $\mathrm{CS}_3$ (see Table~\ref{atab:AZ invariants}). 
Notably, $W_d$ relates to $\mathrm{CS}_d$ through $(-1)^{W_d}=(-1)^{\mathrm{CS}_d}$ for general odd $d$ spatial dimensions~\cite{CTSR-review}. 
Therefore, only the even parts of topological phases in three-dimensional class DIII are sent to the nontrivial phase in three-dimensional class AII. 
This homomorphism is written as ``$\mathbb{Z}^{\checkmark/\times} \xrightarrow{\bmod 2} \mathbb{Z}_2$". One can also prove the presence of such a ``mod 2" map for seven-dimensional class CI in the same manner.

%%%%%%%%%%
\subsection{$\mathbb{Z}_2^{\times} \xrightarrow{0} \mathbb{Z}$}
%\item $\{$``$\mathbb{Z}_2^{\times} \xrightarrow{0} \mathbb{Z}$"$\}$

Because we consider homomorphisms that preserve the group structure, only the possible map from the $\mathbb{Z}_2$ classification to the $\mathbb{Z}$ classification is a zero map. 
In fact, all elements of the $\mathbb{Z}_2$ topological phases are sent to a trivial one in the $\mathbb{Z}$ topological phases. 
We use the symbol ``$\mathbb{Z}_2^{\times} \xrightarrow{0} \mathbb{Z}$" to represent this zero-map homomorphism. 
For instance, in four dimensions, the Fu-Kane invariant $\mathrm{FK}_2$ for class CII does not result in the nontrivial values of the second Chern number $\mathrm{Ch}_2$ for class AII if one ignores both chiral symmetry and particle-hole symmetry.

%%%%%%%%%%
\subsection{$\mathbb{Z}_2^{\times} \xrightarrow{0} \mathbb{Z}_2$}
%\item $\{$``$\mathbb{Z}_2^{\times} \xrightarrow{0} \mathbb{Z}_2$"$\}$

The entries ``$\mathbb{Z}_2^{\times} \xrightarrow{0} \mathbb{Z}_2$" show that all the topological phases in the original AZ classes are mapped to a trivial phase in the corresponding WD classes.
We show this 
using the dimensional reduction of topological phases~\cite{QHZ-08}.
Let us consider three-dimensional class CII as an example. 
This class is characterized by the Chern-Simons invariant $\widetilde{\mathrm{CS}_3}$, while the corresponding WD class (i.e., class AII) is characterized by the other type of Chern-Simons invariant $\mathrm{CS}_3$. 
The dimensional reduction connects the $\mathbb{Z}_2$ descendants to their parent $\mathbb{Z}$ topology in each complex symmetry class. 
For the present case, 
$(-1)^{\widetilde{\mathrm{CS}_3}}=(-1)^{\mathrm{FK}_2}=(-1)^{W_5}$ holds for class CII in $d=3,4,5$. 
Similarly, we have $(-1)^{\mathrm{FK}_1}=(-1)^{\mathrm{CS}_3}=(-1)^{\mathrm{Ch}_2}$ for class AII in $d=2,3,4$; 
see Table~\ref{atab:AZ invariants}. 
As stated in the previous case, $\mathrm{FK}_2$ in four-dimensional class CII always leads to a trivial value of $\mathrm{Ch}_2$ in four-dimensional class AII. 
Combining this with the dimensional reduction in $d=3,4$, we conclude that $\widetilde{\mathrm{CS}_3}$ only relates to trivial $\mathrm{CS}_3$. 
Similarly, one can prove this kind of zero map for other classes.

%%%%%%%%%%
\subsection{$\mathbb{Z}^{\times} \rightarrow 0$,~~$2\mathbb{Z}^{\times} \rightarrow 0$,~~$\mathbb{Z}_{2}^{\times} \rightarrow 0$}
%\item $\{$``$\mathbb{Z}^{\times} \rightarrow 0$", ``$2\mathbb{Z}^{\times} \rightarrow 0$", ``$\mathbb{Z}_{2}^{\times} \rightarrow 0$"$\}$

Since the corresponding WD classes exhibit topologically trivial phases, 
the homomorphisms are just zero maps. 
Here, we do not use the symbol ``$\xrightarrow{0}$" like the previous two cases as the homomorphisms are clearly zero maps.

%%%%%%%%%%
\subsection{$0 \rightarrow 0$,~~$0 \rightarrow \mathbb{Z}$,~~$0 \rightarrow 2\mathbb{Z}$,~~$0 \rightarrow\mathbb{Z}_2$}
%\item $\{$``$0 \rightarrow 0$", ``$0 \rightarrow \mathbb{Z}$", ``$0 \rightarrow 2\mathbb{Z}$", ``$0 \rightarrow\mathbb{Z}_2$"$\}$

Since the original AZ classes are topologically trivial, the homomorphisms are given by zero maps to retain the group structure. 
Here, we do not use the symbol ``$\xrightarrow{0}$" to express the zero maps.

%\end{enumerate}

%%%%%%%%%
\section{$K$-theory perspective}
    \label{sec: K-theory}

In Sec.~\ref{sec: intrinsic}, we classified intrinsic and extrinsic non-Hermitian topology, corresponding to nondetachable and detachable topological boundary states.
In Sec.~\ref{sec: WD}, on the other hand, we classified Wannier localizability of the bulk bands based on the homomorphisms from the AZ classification to the WD classification. 
These two results give the same classification. 
In this section, we discuss why the two approaches coincide with each other from the viewpoint of $K$-theory.
The discussion in this section applies to any magnetic space group symmetry compatible with planar boundary conditions.
We note that the equivalence between the Wannier localizability and detachability of topological boundary states was also argued in Sec.~II\,A of Ref.~\cite{Altland-24}.
Here, we elucidate the underlying mathematical structure based on $K$-theory.

%%%%%%%%%%%%
\subsection{$K$-group}

The degree $+1$ $K$-group $K^1$ is represented by self-adjoint Fredholm operators whose essential spectrum contains both positive and negative values~\cite{AS-69}. 
Gapless Hamiltonians naturally represent such operators realized in topological boundary states, meaning that there exists an energy gap $E_{\rm gap} > 0$ with a finite number of boundary-state spectra in $E \in [-E_{\rm gap}, E_{\rm gap}]$ and an infinite number of bulk spectra outside this range.

Let $\hat H(\bk_\parallel)$ be the infinite-dimensional gapless Hamiltonian on the boundary Brillouin zone $T^{d-1}$ ($\bk_\parallel \in T^{d-1}$), $\mathcal{G}$ be the symmetry group compatible with the presence of the boundary, $\Pi \cong \mathbb{Z}^{d-1}$ be the lattice translation subgroup, and $G = \mathcal{G}/\Pi$ be their quotient group. 
The group $G$ acts on the boundary Brillouin zone as $g: \bk_\parallel \mapsto g\bk_\parallel$. 
We use the homomorphism $\phi, c: G\to \{\pm 1\}$ to specify  
\begin{equation}
    \begin{cases}
        \phi_g = 1: \text{unitary~symmetry}; \\
        \phi_g = -1: \text{antiunitary~symmetry},
    \end{cases}
\end{equation}
and 
\begin{equation}
    \begin{cases}
        c_g = 1: \text{conventional~symmetry}; \\
        c_g = -1: \text{antisymmetry}.
    \end{cases}
\end{equation}
For matrices $M$, we also introduce the notation $M^{\phi_g}$ as $M$ for $\phi_g = 1$ and $M^*$ for $\phi_g = -1$.
Antisymmetry is defined to change the sign of Hamiltonians, prime examples of which include chiral symmetry and particle-hole symmetry.
We summarize the symmetry algebra among symmetry operators in $G$ as 
\begin{align}
    u_g(h\bk_\parallel)u_h(\bk_\parallel)^{\phi_g} = \tau_{g,h}(gh\bk_\parallel) u_{gh}(\bk_\parallel), \quad g,h \in G, 
\end{align}
where $u_g(\bk_\parallel)$ is a unitary operator for $g \in G$. 
Here, $\tau$ is a 2-cocycle $\tau \in Z_{\rm group}^2(G, C(T^{d-1}, U(1)))$, where $C(T^{d-1}, U(1))$ is the group of $U(1)$-valued functions on $T^{d-1}$ with a left $G$-action defined by $(g.\tau)(\bk_\parallel) = \tau(g^{-1}\bk_\parallel)$. 
The twisted equivariant $K$-group ${}^\phi K_G^{(\tau,c)+1}(T^{d-1})$ \cite{FM-13, Gomi-21} classifies topology of $(d-1)$-dimensional gapless Hamiltonians with the following symmetry:
\begin{align}
    u_g(\bk_\parallel) \hat H(\bk_\parallel)^{\phi_g} u_{g}^{\dag}(\bk_\parallel) = c_g \hat H(g\bk_\parallel), \quad 
    g \in G. 
\end{align}

%%%%%%%%%%%%
\subsection{Gapless Hermitian Hamiltonians and point-gapped non-Hermitian Hamiltonians}

First, we relate gapless Hermitian Hamiltonians to point-gapped non-Hermitian Hamiltonians. 
From the gapless Hermitian Hamiltonian $\hat H(\bk_\parallel)$, we obtain the point-gapped non-Hermitian Hamiltonian $H_P(\bk_\parallel)$ by Eq.~(\ref{eq: HP}), with a continuous odd function $\chi(E)$ defined in Eq.~(\ref{eq: chi}).
As a consequence of the symmetries of the gapless Hermitian Hamiltonian $\hat H(\bk_\parallel)$, the obtained point-gapped Hamiltonian $H_P(\bk_\parallel)$ satisfies the following symmetries,
\begin{align}
    u_g(\bk_\parallel) H_P(\bk_\parallel)^{\phi_g} u_g^\dag(\bk_\parallel) = c_g 
    \begin{cases}
        H_P(g\bk_\parallel) & (\phi_g c_g = 1); \\
        H_P^\dag(g\bk_\parallel) & (\phi_g c_g = -1). \\
    \end{cases}
    \label{aeq:sym_Pgap}
\end{align}
In other words, the point-gapped Hamiltonian shares the same unitary symmetries with the original gapless Hamiltonian, while it realizes the other symmetries in a different manner. 
In particular, it hosts the AZ$^{\dagger}$ symmetries, instead of the original AZ symmetries~\cite{KSUS-19}.
Importantly, this map ensures that the $K$-group ${}^\phi K_G^{(\tau,c)+1}(T^{d-1})$ also classifies topology of point-gapped Hamiltonians satisfying symmetry in Eq.~(\ref{aeq:sym_Pgap}). 

We also use another equivalent relation between Hermitian and non-Hermitian topology. 
For the point-gapped Hamiltonian in Eq.~(\ref{eq: HP}), we have the Hermitized Hamiltonian [see also Eq.~(\ref{eq: Hermitization})]
\begin{align}
    \tilde H(\bk_\parallel) = \begin{pmatrix}
        0 & H_P(\bk_\parallel) \\
        H_P^\dag(\bk_\parallel) & 0
    \end{pmatrix}_\sigma, 
\end{align}
which is gapped because of $\tilde{H}(\bk_\parallel)^2=1$.
By design, $\tilde H(\bk_\parallel)$ respects chiral symmetry 
\begin{align}
    \sigma_z \tilde H(\bk_\parallel) \sigma_z^{-1} = - \tilde H(\bk_\parallel).
    \label{aeq:CS_Pgap_doubled}
\end{align}
The symmetry constraints in Eq.~(\ref{aeq:sym_Pgap}) are mapped as 
\begin{align}
    \tilde u_g(\bk_\parallel) \tilde H(\bk_\parallel)^{\phi_g} \tilde u_g^\dag(\bk_\parallel) = c_g \tilde H 
    (g\bk_\parallel), 
    \label{aeq:sym_Pgap_doubled}
\end{align}
with 
\begin{equation}
    \tilde u_g(\bk_\parallel) \coloneqq u_g(\bk_\parallel) \otimes (\sigma_x)^{\frac{1-\phi_g c_g}{2}}, \quad g \in G.
\end{equation}
They satisfy the following algebraic relation with chiral symmetry in Eq.~(\ref{aeq:CS_Pgap_doubled}): 
\begin{align}
(\ii \sigma_z)^2 = -1, \quad \tilde u_g(\bk_\parallel) (\ii \sigma_z)^{\phi_g} = c_g (\ii\sigma_z) \tilde u_g(\bk_\parallel).
    \label{aeq:alg_rel_Pgap_sigmaz}
\end{align}
If chiral symmetry in Eq.~(\ref{aeq:CS_Pgap_doubled}) is absent, the gapped Hamiltonian with $G$ symmetry in Eq.~(\ref{aeq:sym_Pgap_doubled}) gives the degree $0$ $K$-group ${}^\phi K^{(\tau,c)+0}_G(T^{d-1})$. 
Additional chiral symmetry satisfying the algebraic relation in Eq.~(\ref{aeq:alg_rel_Pgap_sigmaz}) increases the degree of the $K$-group by $+1$~\cite{Karoubi}, and thus, the Hermitian Hamiltonian $\tilde{H}(\bk_\parallel)$ gives the same $K$-group ${}^\phi K^{(\tau,c)+1}_G(T^{d-1})$ as the point-gapped Hamiltonian $H_P(\bk_\parallel)$.

%%%%%%%%%%%%
\subsection{Imaginary line gap and detachable topological boundary states}

In the presence of an imaginary line gap, the spectrum of $H_P(\bk_\parallel)$ is defined not to contain ${\rm Im}\,E = 0$.
Additionally, imposing the imaginary line gap is equivalent to imposing the anti-Hermitian condition 
\begin{equation}
H_P^\dag(\bk_\parallel) = -H_P(\bk_\parallel)
    \label{eq: anti-Hermitian}
\end{equation}
on $H_P(\bk_\parallel)$~\cite{KSUS-19}. 
As discussed in Sec.~\ref{sec: intrinsic}, this is also equivalent to imposing chiral symmetry with $\sigma_x$ on the gapped Hermitized Hamiltonian $\tilde H(\bk_\parallel)$ in addition to Eqs.~(\ref{aeq:CS_Pgap_doubled}) and (\ref{aeq:sym_Pgap_doubled}):
\begin{align}
    \sigma_x \tilde H(\bk_\parallel) \sigma_x^{-1} = - \tilde H(\bk_\parallel).
\end{align}
Using this chiral operator $\sigma_x$, we construct a Hamiltonian parameterized by the interval $I = [-1, 1]$ as 
\begin{align}
    \tilde H'(\bk_\parallel, t) = \left( \sin \frac{\pi t}{2} \right) \sigma_x + \left( \cos \frac{\pi t}{2} \right) \tilde H(\bk_\parallel), \quad t \in I.
\end{align}
It should be noted that the symmetry group $G$ acts on the additional parameter space as $c_g t$, 
\begin{align}
    \tilde u_g(\bk_\parallel) \tilde H'(\bk_\parallel, t) \tilde u_g^\dag(\bk_\parallel) = c_g \tilde H'(g\bk_\parallel, c_g t), \quad g \in G.
\end{align}
Therefore, the Hamiltonian $\tilde H'(\bk_\parallel, t)$, and thus the imaginary-line-gapped phase, are classified by the relative $K$-group ${}^\phi K^{(\tau,c)+1}_G(T^{d-1} \times I, T^{d-1} \times \partial I)$, where $G$ acts on $T^{d-1} \times I$ as $g:(\bk_\parallel,t) \mapsto (g\bk_\parallel,c_gt)$, 
and $\partial I = \{-1, 1\}$ is the boundary of the interval. 
From the long exact sequence for the pair $(T^{d-1} \times I, T^{d-1} \times \partial I)$, we obtain the exact sequence 
\begin{align}
    &{}^\phi K^{(\tau,c)+1}_G(T^{d-1} \times I, T^{d-1} \times \partial I) \nonumber \\
    &\qquad\xrightarrow{f^{\rm bdy}_{\rm L_i}}
    {}^\phi K^{(\tau,c)+1}_G(T^{d-1}) \nonumber \\
    &\qquad \xrightarrow{f^{\rm bdy}_0} 
    {}^\phi K^{(\tau,c)+1}_G(T^{d-1} \times \partial I).
\end{align}
Here, we used the isomorphism 
\begin{equation}
    {}^\phi K^{(\tau,c)+1}_G(T^{d-1} \times I) \cong {}^\phi K^{(\tau,c)+1}_G(T^{d-1})
\end{equation}
due to the contractibility of the interval $I$. 
This exactness implies 
\begin{equation}
{\rm im}\ f^{\rm bdy}_{\rm L_i} = {\rm ker}\ f^{\rm bdy}_0.
\end{equation}
Notably, the homomorphism $f^{\rm bdy}_{\rm L_i}$ coincides with the map from imaginary-line-gapped phases to point-gapped phases discussed in Sec.~\ref{sec: intrinsic}~\cite{OKSS-20, Shiozaki}.
If there are no group elements $g \in G$ with $c_g = -1$, the $K$-group ${}^\phi K^{(\tau,c)+1}_G(T^{d-1} \times \partial I)$ reduces to two copies defined on two independent points $\{-1,1\} =\partial I$, leading to:
\begin{align}
    &{}^\phi K^{\tau+1}_G(T^{d-1} \times I, T^{d-1} \times \partial I) \nonumber \\
    &\qquad\xrightarrow{f^{\rm bdy}_{\rm L_i}}
    {}^\phi K^{\tau+1}_G(T^{d-1}) \nonumber \\
    &\qquad \xhookrightarrow{f^{\rm bdy}_0} 
    {}^\phi K^{\tau+1}_G(T^{d-1})^{\oplus 2}.
    \label{eq:ex_seq_w/o_c}
\end{align}
Here, we dropped $c$ owing to $c_g\equiv 1$ from the assumption.
In this case, since $f^{\rm bdy}_0$ sends $x$ to $(x,x)$, we have ${\rm ker}\, f^{\rm bdy}_0=0$ and hence ${\rm im}\, f^{\rm bdy}_{L_{\rm i}}=0$, meaning that there are no point-gapped phases representable by imaginary-line-gapped phases. 
On the other hand, if there is a group element $g \in G$ with $c_g = -1$, $g$ exchanges $t=1$ and $t=-1$, and hence merely relates the $K$-group at $t=1$ with that at $t=-1$. 
Thus, we only need to consider the $K$-group at $t=1$, excluding group elements with $c_g=1$. 
This consideration gives the isomorphism
\begin{equation}
{}^\phi K^{(\tau,c)+1}_G(T^{d-1} \times \partial I) \cong {}^\phi K^{\tau+1}_{G_0}(T^{d-1}), 
\end{equation}
where $G_0 \coloneqq {\rm ker}\ c = \{g \in G \mid c_g = 1\}$ is the subgroup of $G$ excluding antisymmetries.
As a result, we obtain
\begin{align}
    &{}^\phi K^{(\tau,c)+1}_G(T^{d-1} \times I, T^{d-1} \times \partial I) \nonumber \\
    &\qquad \xrightarrow{f^{\rm bdy}_{\rm L_i}}
    {}^\phi K^{(\tau,c)+1}_G(T^{d-1}) \nonumber \\
    &\qquad \xrightarrow{f^{\rm bdy}_0} 
    {}^\phi K^{\tau+1}_{G_0}(T^{d-1}).
    \label{eq:ex_seq_w/_c}
\end{align}
Here, $f^{\rm bdy}_0$ in Eq.~(\ref{eq:ex_seq_w/_c}) can be redefined by forgetting antisymmetries.

To derive Eqs.~(\ref{eq:ex_seq_w/o_c}) and (\ref{eq:ex_seq_w/_c}), we assumed that the point-gapped Hamiltonians $H_P(k_\parallel)$ can be anti-Hermitian. 
Under the same assumption, we can obtain another useful expression for the $K$-group.
For the anti-Hermitian point-gapped Hamiltonian $H_P(\bk_\parallel)$ with Eq.~(\ref{eq: anti-Hermitian}), the symmetry constraints in Eq.~(\ref{aeq:sym_Pgap}) are rewritten as 
\begin{align}
    u_g(\bk_\parallel) (\ii H_P(\bk_\parallel))^{\phi_g} u_g^\dag(\bk_\parallel) = \ii H_P(g\bk_\parallel),\quad g \in G, 
\end{align}
for the Hermitian Hamiltonian $\ii H_P(\bk_\parallel)$. 
Remarkably, antisymmetries with $c_g=-1$ behave as the corresponding symmetries with $c_g=1$. 
Therefore, the imaginary-line-gapped Hamiltonian is classified by the $K$-group ${}^\phi K_G^{(\tau,c_{\rm triv})+0}(T^{d-1})$ with the trivial homomorphism $c=c_{\rm triv}$, i.e., $c_{{\rm triv},g}=1$ for all $g \in G$. 
This is nothing but the Thom isomorphism~\cite{Gomi-21}
\begin{align}
    {}^\phi K^{(\tau,c)+1}_G(T^{d-1} \times I, T^{d-1} \times \partial I) \cong {}^\phi K_G^{(\tau,c_{\rm triv})+0}(T^{d-1}).
\end{align}

In summary, in the absence of antisymmetry ($c_g \equiv 1$), we have the exact sequence 
\begin{align}
    &{}^\phi K^{\tau+0}_G(T^{d-1}) \nonumber \\
    &\qquad \xrightarrow{{\rm im}\, f^{\rm bdy}_{\rm L_i}=0}
    {}^\phi K^{\tau+1}_G(T^{d-1}) \nonumber \\
    &\qquad \xhookrightarrow{f^{\rm bdy}_0} 
    {}^\phi K^{\tau+1}_G(T^{d-1})^{\oplus 2}. 
\end{align}
The triviality of $f^{\rm bdy}_{\rm L_i}$ implies that all the nontrivial gapless boundary Hamiltonians involve spectral flow. 
In the presence of antisymmetries (${}^\exists g\in G, c_g=-1$), on the other hand, we have 
\begin{align}
    &{}^\phi K^{(\tau,c_{\rm triv})+0}_G(T^{d-1}) \nonumber \\
    &\qquad \xrightarrow{f^{\rm bdy}_{\rm L_i}}
    {}^\phi K^{(\tau,c)+1}_G(T^{d-1}) \nonumber \\
    &\qquad \xrightarrow{f^{\rm bdy}_0} 
    {}^\phi K^{\tau+1}_{G_0}(T^{d-1}).
\end{align}
Since the boundary Hamiltonian $\hat H(\bk_\parallel)$ hosts a detachable boundary state when the corresponding point-gapped Hamiltonian may have an imaginary line gap, this equation implies that $\hat H(\bk_\parallel)$ may host a detachable boundary state ($[\hat H(\bk_\parallel)] \in {\rm im}\, f^{\rm bdy}_{\rm L_i}$) if and only if one can open a gap in the boundary state ($[\hat H(\bk_\parallel)] \in {\rm ker}\, f^{\rm bdy}_0$) in $\hat H(\bk_\parallel)$ by ignoring the antisymmetries.

%%%%%%%%%%%%
\subsection{Wannier localizability}

Finally, we clarify the relationship with Wannier localizability. 
It was argued that if the $d$-dimensional bulk admits exponentially localized Wannier states for both occupied and unoccupied states, no spectral flow arises in the boundary Hamiltonian for arbitrary boundary geometry~\cite{Altland-24}. 
Therefore, the existence of spectral flow in the boundary indicates that either occupied or unoccupied states are not Wannier localizable.
To see this in $K$-theory, let us introduce the bulk-to-boundary map
\begin{equation}
f^{\rm BBC}_G: {}^\phi K_G^{(\tau,c)+0}(T^d) \to {}^\phi K^{(\tau,c)+1}_G(T^{d-1}), 
\end{equation}
which sends the bulk Hamiltonian $H(\bk)$, classified by the degree-$0$ $K$-group ${}^\phi K^{(\tau,c)+0}_G(T^d)$, to the boundary Hamiltonian $\hat H(\bk_\parallel)$. 
The bulk-to-boundary map in the absence of antisymmetry
\begin{equation}
f^{\rm BBC}_{G_0}: {}^\phi K_{G_0}^{\tau+0}(T^d) \to {}^\phi K^{\tau+1}_{G_0}(T^{d-1}) 
\end{equation}
is also defined, and forgetting antisymmetries defines the homomorphism 
\begin{equation}
f^{\rm bulk}_0: {}^\phi K_G^{(\tau,c)+0}(T^{d}) \to {}^\phi K_{G_0}^{\tau+0}(T^{d}) 
\end{equation}
for the bulk. 
Then, we obtain the commutative diagram in Eq.~(\ref{eq:exact_seq}) in which the right vertical line is exact.
Given a bulk Hamiltonian $[H(\bk)] \in {}^\phi K_G^{(\tau,c)+0}(T^d)$, the existence of spectral flow at the boundary means 
\begin{equation}
f^{\rm BBC}_G([H(\bk)]) \notin {\rm im}\, f^{\rm bdy}_{L_{\rm i}} = {\rm ker}\, f^{\rm bdy}_0.
\end{equation}
The commutativity of the diagram in Eq.~(\ref{eq:exact_seq}) implies 
\begin{equation}
f^{\rm BBC}_{G_0} \circ f^{\rm bulk}_0([H(\bk)]) \neq 0 \in {}^\phi K_{G_0}^{\tau+1}(T^{d-1}). 
\end{equation}
This further means that the bulk topological invariant without antisymmetry, which detects the element of the $K$-group $K_G^{\tau+0}(T^d)$ signaling the boundary spectral flow, should provide the Wannier obstruction.

%%%%% Discussions %%%%%
\section{Conclusion}
    \label{sec: conclusion}

In this work, we elucidate the relationship between Wannier localizability and detachability of topological boundary states.
From the boundary perspective, we classify intrinsic and extrinsic non-Hermitian topology, underlying nondetachable and detachable topological boundary states, respectively.
From the bulk perspective, by contrast, we classify Wannier localizability by the homomorphisms of topological phases from the tenfold AZ symmetry classes to the threefold WD symmetry classes.
We clarify the agreement of the two classification from the boundary and bulk perspectives on the basis of $K$-theory.
Our work, combined with our accompanying Letter~\cite{NSSSK-24}, provides a fundamental understanding of Wannier localizability and detachable boundary states, as well as the bulk-boundary correspondence, in topological materials.
In passing, while we focus on stable topological insulators in this work, Hopf insulators likewise exhibit topological boundary states detachable from the bulk bands~\cite{Moore-08, Alexandradinata-21}.
It is worthwhile to investigate these detachable boundary states within our framework.

\medskip
%%%%% Note %%%%%
{\it Note added}.---After the completion of this work, we became aware of a recent related work~\cite{Lapierre-24}.

\medskip
%%%%% Acknowledgement %%%%%
\begingroup
\renewcommand{\addcontentsline}[3]{}% Remove functionality of \addcontentsline
\begin{acknowledgments}
We thank Akira Furusaki, Shingo Kobayashi, and Shinsei Ryu for helpful discussion.
We appreciate the long-term workshops ``Recent Developments and Challenges in Topological Phases" (YITP-T-24-03) and ``Dynamics Days Asia Pacific 13 (DDAP13) / YKIS2024" held at Yukawa Institute for Theoretical Physics (YITP), Kyoto University.
K. Shiozaki, D.N., K. Shimomura, and M.S. are supported by JST CREST Grant No.~JPMJCR19T2. 
K. Shiozaki is supported by JSPS KAKENHI Grant Nos.~JP22H05118 and JP23H01097. 
D.N. is supported by JSPS KAKENHI Grant No.~JP24K22857.
K. Shimomura is supported by JST SPRING, Grant No.~JPMJSP2110 and JSPS KAKENHI Grant No.~JP25KJ1632.
M.S. is supported by JSPS KAKENHI Grants No.~JP24K00569 and No.~JP25H01250. 
K.K. is supported by MEXT KAKENHI Grant-in-Aid for Transformative Research Areas A ``Extreme Universe" No.~JP24H00945.
\end{acknowledgments}
\endgroup

\let\oldaddcontentsline\addcontentsline% Store \addcontentsline
\renewcommand{\addcontentsline}[3]{}% Make \addcontentsline a no-op
\bibliography{NH_top.bib}
\let\addcontentsline\oldaddcontentsline% Restore \addcontentsline

\end{document}